\documentclass[prl]{revtex4}
\usepackage{graphicx}
\usepackage{verbatim}
\usepackage{color}
\usepackage{bm}
\usepackage{pbox}
\usepackage{longtable}
\usepackage{graphics}
\usepackage{amsmath}
\usepackage{bm}
\usepackage{amsfonts}
\usepackage{amssymb}
\usepackage[T1]{fontenc}
\usepackage[latin9]{inputenc}
\usepackage{esint}
\usepackage{slashbox}

\begin{document}

\title{Game-theoretic perspective of Ping-Pong Protocol}
\author{Hargeet Kaur} 
\affiliation{Indian Institute of Technology Jodhpur, Jodhpur, Rajasthan, India}
\author{Atul Kumar}
\email{atulk@iitj.ac.in}
\affiliation{Indian Institute of Technology Jodhpur, Jodhpur, Rajasthan, India}
\date{\today}

\begin{abstract}
We analyse Ping-Pong protocol from the point of view of a game with quantum strategies. The analysis helps us in understanding the different strategies of a sender and an eavesdropper to gain the maximum payoff in the game. The study presented here characterizes strategies that lead to different Nash equilibriums. We further demonstrate the condition for Pareto optimality depending on the parameters used in the game. Moreover, we also analysed LM05 protocol and compared it with PP protocol from the point of view of a generic two-way QKD game with or without entanglement. Our results provide a deeper understanding of general two-way QKD protocols in terms of the security and payoffs of different stakeholders in the protocol.  
\end{abstract}
\maketitle

\section{Introduction}
A game is a competitive activity among more than one rational players, where there are a set of rules and conditions of win and loss. Each player makes a strategic move depending on certain background details such as knowledge about other players, knowledge about allowed moves, and how different moves will lead to varying outcomes of the game. Every move that a player takes correspond to a {\em{Payoff}} - payoffs are numbers which represent the benefit that each player gets by their respective moves. It quantifies the utility or the desirability of each player to perform a particular strategy. For a finite game, John F. Nash \cite{Nash1,Nash2} described a stable point -Nash Equilibrium- that is formulated by those strategy sets where no player gets an incentive by unilaterally changing her/his strategy. A strategy set of a game is Pareto efficient (or Pareto optimal) if there is no other strategy set that makes atleast one player better off without making any other player worse off. \par
The detailed analysis of such a strategic decision-making in any competitive situation is inherent in game theory \cite{Von Neumann}. Since its inception, the theory has found applications in diverse academic spaces such as economics, political science, biology, computer science, physics etc  \cite{eco1,eco2,pol1,bio1,bio2,cse1,cse2,cse3,cse4,physics1}. With the advent of quantum information and computation, the quest to analyse classical game theory in quantum realm became central to the foundations of quantum mechanics \cite{Wiesner}. Meyer \cite{Meyer} and Eisert \cite{Eisert} independently put classical game theory in the context of quantum strategies. The central idea to introduce quantum strategies in comparison to classical strategies is to achieve a better payoff in the game. For example, Meyer demonstrated that a quantum player always outperforms a classical player in a Penny flip game and Eisert described how quantum strategies help players to avoid the original dilemma present in the classical Prisoners' dilemma (PD). On the experimental front, the quantum version of the PD game is also realized using a NMR quantum computer \cite{Du}. Moreover, Vaidman \cite{Vaidman} illustrated a simple game in which players always win the game when they share a GHZ state in advance, in comparison to three classical players where the probability of winning the game is always probabilistic. Quantum strategies are also utilized to introduce the elements of fairness in remote gambling \cite{Goldenberg}, and in designing algorithms for implementing quantum auctions which offer many security advantages \cite{Patel}. Flitney and Abbott \cite{Flitney} have analysed quantum versions of Parrondo's games. Such analysis not only helps to design secure networks that lead to identification of new quantum algorithms, but also provide a completely different dimension to characterize a game or a protocol. Furthermore, eavesdropping \cite{Ekert,Gisin} and optimal cloning \cite{Werner} can also be visualized as games between players. \par
Applications of Quantum game theory are gaining importance, since lesser bits are used to play quantum games \cite{Patel}. Quantum game theory also becomes important as one can represent quantum communication protocols, and algorithms, in terms of games between quantum and classical players \cite{Iqbal}. A quantum game differs from a classical game in three principal ways. Firstly, the states employed in a quantum game can be visualized as a quantum superposition of two or more base states. Moreover, the players must initially share entangled states, and can choose to perform any superposition of strategies on the initial state. In this article, we revisit Ping-Pong (PP) protocol \cite{Bostrom1} from the perspective of a game between the sender and the eavesdropper. We have limited our analysis to a classical selection of strategies by the sender and the eavesdropper. We have not included a superposition of strategies for the players, and hence we have given a classical game-theoretic picture of PP protocol. \par
Our results demonstrate how pure strategy Nash equilibrium changes depending on the payoffs of the two players. From the point of view of Alice, the Nash equilibrium would illustrate a strategy that Alice must use for encoding information and from Eve's point of view, the Nash equilibrium will be the most information gaining attack. We further analyse the strategy that a sender must employ to ensure minimum payoff to an eavesdropper. On the other hand, we also describe the best strategy for the sender and eavesdropper to settle for a Pareto optimal Nash equilibrium, only from the perspective of a general game and not from the perspective of a secure protocol, depending on certain parameters which play essential role in the protocol. In addition, we also study another two-way QKD protocol, i.e., LM05 protocol from the perspective of a game and compare it with PP game to analyse general payoffs of the players in a game with or without entanglement. We found that depending on the protocol or game (with or without  entanglement) and weights involved in the payoff term, different strategies of players may lead to different Nash equilibriums. The perspective used here, therefore, provides a deeper understanding of the protocol in terms of security, eavesdropping and importance of different parameters which are part of the protocol.   \par

\section{Ping-Pong protocol as a Game with Quantum strategies}
Any activity which involves dealing with competitive situations can be a game. For example, any communication protocol where a sender (Alice) wants to securely transfer information to a receiver (Bob) can be a game between Alice and an eavesdropper (Eve) who does not want Alice to successfully complete her job. In order to win the game, Eve may try to gain the secret information and/or modify the information that Alice wants to send to Bob. On the contrary, Alice will try to employ a strategy such that Eve  is unable to intervene in any way. This will result in a game for different strategies of Alice and Eve in a protocol. Therefore, game theory can be used for easy and detailed understanding of many communication protocols, e.g., key distribution protocols.   \par
Quantum key distribution (QKD) protocols are proposed with single and entangled quantum systems. For example BB84 \cite{Bennett} is an example of one-way single-photon QKD protocol and PP Protocol is an example of two-way QKD protocol based on entangled photons. BB84 protocol has been earlier studied well within the set-up of a game \cite{Houshmand}. In this article, we will study PP protocol to analyse the strategies of a sender and an eavesdropper. PP protocol uses entanglement to allow asymptotically secure key distribution and quasi-secure direct communication. Here, Eve has access to one of the photons at two different stages; once during entanglement distribution and once after the encoding of secret information. Therefore, there are chances that the Eavesdropper may try to gain some information communicated from a sender to the intended receiver. For practical purposes, Ostermeyer and Walenta \cite{Ostermeyer} have also proposed the experimental realization of PP protocol. In order to facilitate our analysis, we first describe the PP protocol to understand the different aspects of the protocol. \par      
In the original protocol, Bob prepares a Bell state $\vert{\psi^{+}}\rangle_{AB}=\dfrac{1}{\sqrt{2}}(\vert{01}\rangle+\vert{10}\rangle)_{AB}$ and sends the particle A (travel photon) to Alice and keeps particle B (home photon) with himself. Alice randomly operates between control and message modes. In control mode, she measures the travel photon in computational basis and announces the result to Bob, who then measures his photon in the same basis. If the measurement outcomes of Alice and Bob are correlated as in 
 $\vert{\psi^{+}}\rangle_{AB}$, then they proceed with the communication; else an eavesdropper is detected and the protocol is aborted. In message mode, Alice performs unitary operation $I$ or $\sigma_{z}$ on the travel photon to encode 0 or 1, respectively. After encoding, she sends the travel photon back to Bob, who performs a Bell state measurement on the joint state of two photons. The measurement outcome  
$\vert{\psi^{+}}\rangle$ indicates that Alice performed $I$ operation and the measurement outcome
$\vert{\psi^{-}}\rangle$ indicates that Alice performed $\sigma_{z}$ operation. Therefore depending on the measurement outcomes, Bob decodes the one-bit information communicated by Alice. \par
In the entire protocol, there are two instances where the travel photon could be attacked by Eve. First, when it was sent from Bob to Alice for entanglement distribution and second, when it was sent from Alice to Bob after encoding. Since the eavesdropper does not know in which turn Alice will operate in the control mode, she will attack the travel photon each time it is sent, irrespective of the control or message mode. For this, Eve  introduces auxiliary photons $\vert{v}\rangle_{x}\vert{0}\rangle_{y}$ to the shared state during entanglement distribution, where $\vert{v}\rangle$ denotes a vacuum state. Eve then manipulates the joint state of travel and auxiliary photons by performing unitary operation $Q$. Wojcik \cite{Wojcik} described an attack operation by Eve on PP protocol, wherein Eve gets detected with $50\%$ probability in the control mode. In the process Eve gains some information, thereby reducing the mutual information between the sender and the receiver. In fact a symmetrization procedure to this attack further reduces the amount of mutual information between Alice and Bob. Wojcik's eavesdropping attack was later improved to reduce the induced channel loss in control mode to $25\%$ \cite{Zhang}. Moreover, a \textit{denial-of-service} (DoS) attack and an improvement to increase the capacity of PP protocol was also studied \cite{Cai1,CaiLi1}. He further showed how imperfect implementation of the protocol could be exploited by an invisible photon eavesdropping with zero detection risk \cite{Cai2}. The security of the PPP was however, reviewed in light of several attacks \cite{Bostrom2}.  \par 
Interestingly, Pavicic \cite{Pavicic} introduced a slightly different attack operation on PP protocol where the probability of detection of Eve in the control mode gets reduced to zero, but in the process Eve also does not get any information. In this article, we have analysed a situation wherein an eavesdropper refrains from performing any attack and simply performs an identity operator. By performing such an operator on PP protocol, Eve remains undetected in the control mode but does not gain any information. The advantage with our attack is in terms of resources, i.e., Eve uses no gate in comparison to Pavicic's attack. For our purpose, we assume that Eve introduces auxiliary photons $\vert{v}\rangle_{x}\vert{0}\rangle_{y}$ to the shared state. Eve's attack operations on the travel and auxiliary photons are represented as 
\begin{eqnarray}
Q=I
\end{eqnarray}
\par Based on the above discussion, we now proceed to analyse the PP protocol from the perspective of a game. The different strategies of Alice and Eve will be used to formulate a game between them. For example, the rules of game are designed in a way such that Alice's payoff increases when she sends more information to Bob and lesser information is leaked out to Eve. Thus, the mutual information shared between Alice and Bob plays a positive role in the payoff of Alice and the mutual information shared between Alice and Eve, and Bob and Eve plays a negative role in the payoff of Alice. Also, if Alice is able to detect the presence of Eve, it will help her getting a better payoff in this competitive situation. On the other hand, Eve's payoff increases by an increase in the amount of information that Eve gains from Alice and Bob, and decreases by an increase in mutual information between Alice and Bob. Furthermore, Eve will loose points if she gets detected, and therefore the probability of Eve not being detected increases the payoff of Eve. In addition, Eve also applies some gates to gain information from Alice and Bob. More the number of gates, more will be the overhead of Eve and this will play a negative role in the payoff of Eve. Hence, the payoff can be designed according to the various situations of benefit of players that we wish to analyse. Summing up all the factors described above, we can formulate the payoff of Alice as
\begin{eqnarray}
P_{A}=w_{a}I(A:B)-w_{b}[I(A:E)+I(B:E)]+w_{c}p_{d}
\end{eqnarray}
and the payoff of Eve as
\begin{eqnarray}
P_{E} &=& w_{d}[I(A:E)+I(B:E)]-w_{e}I(A:B) + w_{f}[1-p_{d}]-w_{1}n_{1}-w_{2}n_{2}-w_{3}n_{3}
\end{eqnarray}
where $w_{a}, w_{b}, w_{c}, w_{d}, w_{e}, w_{f}, w_{1}, w_{2}, w_{3}$ are positive real numbers and considered as weights attached to each quantity in the payoff, $I(A:B)$ is the mutual information between Alice and Bob, $I(A:E)$ is the mutual information between Alice and Eve, $I(B:E)$ is the mutual information between Bob and Eve, $p_{d}$ is the probability of detection of Eve, $n_1$ is the number of two qubit gates, $n_2$ is the number of single qubit gates and $n_3$ is the number of beam splitters in the attack operation of Eve. Since the payoffs of Alice and Eve depend on different weight values, this kind of game is not a zero-sum game, i.e., there is no complete win or a complete loss situation for any player. The players are always benefited to some degree quantified by $P_A$ and $P_E$ in Eq. (2) and Eq. (3), respectively. \par
We have designed the payoff of Alice and Eve to study a general scenario, and therefore we have kept an account of all possible terms that may contribute towards the payoff of Alice and Eve. In order to study specific scenarios, we can choose different values of weights in the payoff to study the game. For example, 
we can study eavesdropping attacks without considering \textit{denial-of-service} type attacks, by considering weights $w_b$ and $w_e$ to be zero. Similarly, if we wish to analyse a PP game where the eavesdropper has unlimited power constrained only by the laws of physics, i.e., Eve is not bound by the cost of resources then we can assume weights $w_1$, $w_2$, and $w_3$ to be zero. We will analyse the specific scenarios as explained above later. For studying other special cases of the game, one can assign varying values to the weights in the payoff. \par
From the point of view of strategies adopted by Alice and Eve, we consider four different attack operations as the strategy of Eve, namely Wojcik's original attack \cite{Wojcik}, Wojcik's symmetrized attack \cite{Wojcik}, Pavicic's attack \cite{Pavicic} and no attack or an identity operator as in Eq. (1). In comparison to Eve's strategies we consider two different strategies of Alice for encoding one bit information. Since we are analysing the PP protocol as a game, it becomes important to consider more than one allowed strategy (move) for Alice, for a comparative analysis. Therefore, we considered phase flip encoding and bit flip encoding as the two different strategy sets of Alice. Phase flip encoding can be implemented by performing $I$ on the travel photon to send $0$ and $\sigma_z$ on the travel photon to send $1$. On the other hand, bit flip encoding could be a slight variant of the above encoding strategy, i.e., performing $I$ on the travel photon to send $0$ and $\sigma_x$ on the travel photon to send $1$. For the two strategies of Alice, we have modified the protocol slightly to make sure that each player remains unaware of the other player's move. Bob decodes information about the encoding scheme after the travel photon finally reaches him. Bob announce the receipt of the travel photon, after which Alice announces her strategy $A_1$ or $A_2$. Thus, Eve may come to know about the encoding scheme of Alice, after she is done with her move and cannot apply additional operations or moves. This way, Eve can take her move (eavesdropping operation) without knowing Alice's move (encoding operation).

\section{Similarity of Ping-Pong protocol to the Messenger Game}
The PP protocol, as a game, holds similarity to a modified form of messenger game/ whisper down the lane game. In this game, there are multiple players (let, $n+2$) sitting in a queue. The first player (sender or Alice) whispers a message to the ear of the next person through a line of ($n$) players until the last player (receiver or Bob) receives the final message. The aim of the game is the same as that of any communication protocol that the message should reach the receiver unaltered. We can slightly modify this game by assuming that there is a mischievous player (Eve) among ($n$) players who gets to hear the actual message the sender intends to send, but modifies the intended message and passes on the modified message to her next neighbour, as her aim is not to let Bob know the actual message. \par
As discussed above, Alice randomly chooses to operate in control mode or in message mode. The message mode is similar to the usual messenger game, where Alice, i.e., the first person whispers the desired message through the line of players until Bob, i.e., the last player receives the message and acknowledges the receipt of the message. In control mode, Alice plays the usual messenger game, but sends a dummy message. When Bob announces the receipt of message while in control mode, Alice randomly asks the $i^{th}$ person in the queue about the message he/she was asked to transfer to his/her neighbour in line. If the $i^{th}$ person says the same message which Alice had sent, the segment of doubt (where an Eve may be present) for the next control mode, reduces to players between $i^{th}$ to $n^{th}$ position. On the other hand, if the $i^{th}$ person says any message which is different from what Alice had sent, the segment of doubt for the next control mode, reduces to players between $1^{st}$ to $i^{th}$ position.  This way, after a finite number of control runs for the game, if "$d$" is the number of players in the segment of doubt, then the probability of detection of Eve will be $1/d$. Now, since Eve knows that there can be random control runs between message modes, Eve randomly chooses not to modify the transferred message, so as to avoid being caught during the control mode. This random guessing by Eve could correspond to the overhead of Eve in the form of single and double qubit gates, and polarization beam splitters in the PP protocol. Thus, the payoffs of the players in the PP game as designed in Eq. (2) and Eq. (3) hold similarity to the payoffs of Alice and Eve for the above described modified messenger game. Therefore, one can easily relate and understand PP protocol as a game when visualized as a more familiar messenger game.

\section{Analysis of different Strategies for the Game}
We consider two different strategies for Alice, namely $A_1= encoding Scheme(0:I,1:\sigma_{z})$ and $A_2= encoding Scheme(0:I,1:\sigma_{x})$, and four different strategies for Eve, namely $E_1$ - Wojciks's attack, $E_2$ - symmetrized Wojcik's attack, $E_3$ - Pavicic's attack, and $E_4$ - no attack. For each strategy $A_i$ and $E_j$ ($i\in\lbrace1,2\rbrace$,$j\in\lbrace1,2,3,4\rbrace$), the payoff of Alice and Eve can be calculated from Eq. (2) and Eq. (3), respectively. 

\vspace*{-5mm}
\begin{table}[h!]
  \centering
  \caption{Payoffs of Alice in the PP game}
  \label{tab:table1}
  \begin{tabular}{|c|c|c|c|c|}
  \hline
    \backslashbox{Alice}{Eve} & $E_1$ & $E_2$ & $E_3$ & $E_4$\\
    \hline
    $A_1$ & $0.311w_{a}-0.385w_{b}+0.5w_{c}$ & $0.188w_{a}-0.377w_{b}+0.5w_{c}$ & $w_{a}$ & $w_{a}$ \\
    \hline
    $A_2$ & $0.311w_{a}-0.86w_{b}+0.5w_{c}$ & $0.423w_{a}-0.768w_{b}+0.5w_{c}$ & $w_{a}-2w_{b}$ & $w_{a}$ \\
    \hline
  \end{tabular}
\end{table}

\vspace*{-3mm}
\begin{table}[h!]
  \centering
  \caption{Payoffs of Eve in the PP game}
  \label{tab:table2}
  \renewcommand{\arraystretch}{2}
  \begin{tabular}{|c|c|c|c|c|}   \hline
  \backslashbox{Alice}{Eve} & $E_1$ & $E_2$ & $E_3$ & $E_4$ \\  [1.0 ex]   \hline
    $A_1$ & \pbox{20cm} {$0.385w_{d}-0.311w_{e}+0.5w_{f}$ \nonumber \\ $-10w_{1}-4w_{2}-2w_{3}$} & \pbox{20cm} {$0.377w_{d}-0.188w_{e}+0.5w_{f}$ \\ $-10.5w_{1}-5.5w_{2}-2w_{3}$} & \pbox{20cm} {$-w_{e}+w_{f}-8w_{1}$ \\ $-4w_{2}-2w_{3}$} & \pbox{20cm} {$-w_{e}+w_{f}$} \\ [2.0 ex]     \hline
    $A_2$ & \pbox{20cm} {$0.86w_{d}-0.311w_{e}+0.5w_{f}$ \\ $-10w_{1}-4w_{2}-2w_{3}$} & \pbox{20cm} {$0.768w_{d}-0.423w_{e}+0.5w_{f}$ \\ $-10.5w_{1}-5.5w_{2}-2w_{3}$} & \pbox{20cm} {$2w_{d}-w_{e}+w_{f}-8w_{1}$ \\ $-4w_{2}-2w_{3}$} & \pbox{20cm} {$-w_{e}+w_{f}$} \\ [2.0 ex]    \hline
  \end{tabular}
\end{table}

Table I and Table II summarize the payoffs of Alice and Eve, respectively based on the strategies they opt for in the game. One can easily observe from these tables that the eavesdropper always gets a lesser payoff for all her strategies whenever Alice performs the strategy $A_1$. Hence, from the perspective of protocol's security, Alice may prefer to opt for the strategy $A_1$. However, from the perspective of a game between two players, we further discuss some of the outcomes of different strategies opted by Alice and Eve, such that
\begin{itemize}
\item If Eve performs $E_1$ or $E_3$ then Alice is better off by performing $A_1$ because 
\begin{eqnarray}
&& 0.311w_{a}-0.385w_{b}+0.5w_{c}\geq 0.311w_{a}-0.86w_{b}+0.5w_{c}, \,\, and \, \,  \nonumber \\
&& w_{a}\geq w_{a}-2w_{b}
\end{eqnarray}
\item If Eve performs $E_4$, then Alice gets an equal payoff by performing either $A_1$ or $A_2$ 
\item If Eve performs $E_2$, then Alice can be better off by performing $A_1$ or $A_2$ depending on the values of $w_{a}$ and $w_{b}$ 
\begin{eqnarray}
A_{1}:   \, \, 0.188w_{a}-0.377w_{b}+0.5w_{c}\geq 0.423w_{a}-0.768w_{b}+0.5w_{c} &\Rightarrow & w_{b}\geq 0.601w_{a}, \,\, and \nonumber \\
A_{2}:  \, \, 0.188w_{a}-0.377w_{b}+0.5w_{c}\leq 0.423w_{a}-0.768w_{b}+0.5w_{c} &\Rightarrow & w_{b}\leq 0.601w_{a}
\end{eqnarray}
\item Assuming that $0.123w_{e}-0.008w_{d}\leq 0.5w_{1}+1.5w_{2}$, if Alice performs $A_1$ then Eve gets lesser payoff by performing $E_2$ and $E_3$ in comparison to performing $E_1$ or $E_4$. Eve can opt the strategy $E_1$ or $E_4$ depending on the value of weights $w_d$, $w_e$, $w_f$, $w_1$, $w_2$ and $w_3$, i.e., if     
\begin{eqnarray}
&& 0.385w_{d}-0.311w_{e}+0.5w_{f}-10w_{1}-4w_{2}-2w_{3}\geq -w_{e}+w_{f} \nonumber \\ 
&\Rightarrow & 0.385w_{d}+0.689w_{e}\geq 0.5w_{f}+10w_{1}+4w_{2}+2w_{3} 
\end{eqnarray}
then Eve prefers $E_1$, else she prefers $E_4$. 
\item For $0.123w_{e}-0.008w_{d}\geq 0.5w_{1}+1.5w_{2}$ and Alice performing $A_1$, Eve gets higher payoff by performing $E_2$ or $E_4$ strategy. Similar to the above case, if 
\begin{eqnarray}
&& 0.3775w_{d}-0.188w_{e}+0.5w_{f}-10.5w_{1}-5.5w_{2}-2w_{3}\geq -w_{e}+w_{f} \nonumber \\  
&\Rightarrow & 0.377w_{d}+0.812w_{e}\geq 0.5w_{f}+10.5w_{1}+5.5w_{2}+2w_{3}
\end{eqnarray}
then Eve prefers $E_2$, else she prefers $E_4$.
\item If $w_{d}\leq 4w_{1}+2w_{2}+w_{3}$ and Alice performs $A_2$, then Eve gets better payoff by performing $E_1$ or $E_4$. The highest payoff strategy between $E_1$ and $E_4$ depends on the value of the weights $w_d$, $w_e$, $w_f$, $w_1$, $w_2$ and $w_3$, such that if
\begin{eqnarray}
&& 0.86w_{d}-0.311w_{e}+0.5w_{f}-10w_{1}-4w_{2}-2w_{3}\geq -w_{e}+w_{f} \nonumber \\
&\Rightarrow & 0.86w_{d}+0.689w_{e} \geq 0.5w_{f}+10w_{1}+4w_{2}+2w_{3}
\end{eqnarray}
\begin{table}[h!]
  \centering
  \caption{Conditions for $(A_{i}, E_{j})$ to be a Nash equilibrium}
  \label{tab:table3}
  \renewcommand{\arraystretch}{2}
  \begin{tabular}{|c|c|}   \hline
   Nash Equilibrium & Conditions \\  [1.0 ex]   \hline
    $(A_1,E_1)$ & \pbox{20cm} {$0.123w_{e}-0.008w_{d} \leq 0.5w_{1}+1.5w_{2}$, and \nonumber \\ $0.385w_{d}+0.689w_{e} \geq 0.5w_{f}+10w_{1}+4w_{2}+2w_{3}$} \\ [2.0 ex] \hline
    $(A_1,E_2)$ & \pbox{20cm} {$w_{b} \geq 0.601w_{a}$, \nonumber \\ $0.123w_{e}-0.008w_{d} \geq 0.5w_{1}+1.5w_{2}$, and \nonumber \\ $0.377w_{d}+0.812w_{e} \geq 0.5w_{f}+10.5w_{1}+5.5w_{2}+2w_{3}$} \\ [2.0 ex] \hline
    $(A_1,E_4)$ & \pbox{20cm} {$0.123w_{e}-0.008w_{d} \leq 0.5w_{1}+1.5w_{2}$, and \nonumber \\ $0.385w_{d}+0.689w_{e} \leq 0.5w_{f}+10w_{1}+4w_{2}+2w_{3}$} \\ [2.0 ex] \hline
    $(A_2,E_4)$ & \pbox{20cm} {$w_{d} \leq 4w_{1}+2w_{2}+w_{3}$, and \nonumber \\ $0.86w_{d}+0.689w_{e} \leq 0.5w_{f}+10w_{1}+4w_{2}+2w_{3}$} \\ [2.0 ex] \hline 
  \end{tabular}
\end{table} 
then Eve is better off by performing $E_1$, else she performs $E_4$. 
\item Similarly for $w_{d}\geq 4w_{1}+2w_{2}+w_{3}$ and Alice's strategy $A_2$, Eve gets higher payoff by performing $E_1$ or $E_3$, such that if 
\begin{eqnarray}
&& 0.86w_{d}-0.311w_{e}+0.5w_{f}-10w_{1}-4w_{2}-2w_{3}\geq 2w_{d}-w_{e}+w_{f}-8w_{1}-4w_{2}-2w_{3} \nonumber \\
&\Rightarrow & 0.689w_{e}-1.14w_{d}\geq 0.5w_{f}+2w_{1}
\end{eqnarray}
then Eve prefers $E_1$, else she prefers $E_3$. 
\end{itemize}
\vspace*{5mm}
Summing up the above discussion, we can easily conclude that the Nash equilibrium of the generic game is either ($A_1$,$E_1$), ($A_1$,$E_2$), ($A_1$,$E_4$) or ($A_2$,$E_4$) depending on the values of the weights as indicated in Table III. The Nash equilibrium for specific cases such as eavesdropping or Eve equipped with unlimited resources will differ from the ones represented in Table III as we will demonstrate later. For the general case and to simplify our analysis, we consider all the weights attached to the mutual information terms and all the weights attached to the probability terms to be independently equal to each other, i.e.,
\begin{eqnarray}
&& w_{a}=w_{b}=w_{d}=w_{e}=w_{I}, \, \, \, and \nonumber\\
&& w_{c}=w_{f}=w_{P}
\end{eqnarray}
Table III together with Eq. (12) illustrates that for $(A_1,E_1)$ or $(A_1,E_2)$ to be the Nash equilibrium of the game, $w_{I}$ must have a very high value in comparison to the values of $w_{P}$, $w_{1}$, and $w_{2}$ which may not be a feasible choice under real situations. Therefore, ($A_1$,$E_4$) and/or ($A_2$,$E_4$) become the Nash equilibrium of the one bit ping-pong game. 
\vspace*{-2mm}
\begin{table}[h!]
  \centering
  \caption{Payoffs of Alice in the PP game}
  \label{tab:table4}
  \begin{tabular}{|c|c|c|c|c|}
  \hline
    \backslashbox{Alice}{Eve} & $E_1$ & $E_2$ & $E_3$ & $E_4$\\
    \hline
    $A_1$ & $-0.074w_{I}+0.5w_{P}$ & $-0.189w_{I}+0.5w_{P}$ & $w_{I}$ & $w_{I}$ \\
    \hline
    $A_2$ & $-0.549w_{I}+0.5w_{P}$ & $-0.345w_{I}+0.5w_{P}$ & $-w_{I}$ & $w_{I}$ \\
    \hline
  \end{tabular}
\end{table}
\vspace*{2mm}
Interestingly, $E_4$ is Eve's strategy where she does not get detected, gains no information, and uses no gates. Therefore, for a particular situation where Eve may need to optimize her resources, even the attack operation $E_4$ (equivalent to Eve doing nothing) becomes useful in a game situation. The attack operation $E_4$, however, may not be relevant for situations where Eve is equipped with unlimited resources. From Eq. (12), one can show that Table I and Table II reduce to Table IV and  Table V, respectively. \\
\vspace*{-2mm}
\begin{table}[h!]
  \centering
  \caption{Payoffs of Eve in the PP game}
  \label{tab:table5}
  \renewcommand{\arraystretch}{2}
  \begin{tabular}{|c|c|c|c|c|}   \hline
    \backslashbox{Alice}{Eve} & $E_1$ & $E_2$ & $E_3$ & $E_4$ \\  [1.0 ex]   \hline
    $A_1$ & \pbox{20cm} {$0.074w_{I}+0.5w_{P}-10w_{1}$ \nonumber \\ $-4w_{2}-2w_{3}$} & \pbox{20cm} {$0.189w_{I}+0.5w_{P}-10.5w_{1}$ \\ $-5.5w_{2}-2w_{3}$} & \pbox{20cm} {$-w_{I}+w_{P}-8w_{1}$ \\ $-4w_{2}-2w_{3}$} & \pbox{20cm} {$-w_{I}+w_{P}$} \\ [2.0 ex]     \hline
    $A_2$ & \pbox{20cm} {$0.549w_{I}+0.5w_{P}-10w_{1}$ \\ $-4w_{2}-2w_{3}$} & \pbox{20cm} {$0.345w_{I}+0.5w_{P}-10.5w_{1}$ \\ $-5.5w_{2}-2w_{3}$} & \pbox{20cm} {$w_{I}+w_{P}-8w_{1}$ \\ $-4w_{2}-2w_{3}$} & \pbox{20cm} {$-w_{I}+w_{P}$} \\ [2.0 ex]    \hline
  \end{tabular}
\end{table}
\vspace*{2mm}
Although from the perspective of PP protocol, the players will never strive for a Pareto optimal Nash equilibrium. However, when we visualize a communication protocol as a game, the prominence of a Pareto optimal Nash equilibrium comes into picture. In a game perspective, there is either win or lose situation. Whereas in a communication protocol, there can be many aspects, like secure transmission of information, control runs for any third party detection, etc. Therefore, whenever we switch from a protocol to its game counterpart, it becomes essential to analyse the Nash equilibrium of the game. The Nash equilibrium may not be the condition that the players would opt for in a secure protocol. But, in the game theoretic view, the greed of the players for achieving maximum possible payoff drives them to Nash equilibrium. We, therefore, analyse the Pareto optimal strategy for both the players. For the PP game described above $(A_1,E_4)$ and  $(A_2,E_4)$ will be the Pareto optimal Nash equilibrium of the game if $w_I$ is the highest payoff of Alice in Table IV, and $-w_I+w_P$ is the highest payoff of Eve in Table V. Since a Pareto optimal strategy is the one in which players do not get a higher incentive by changing their strategies, $(A_1,E_4)$ and $(A_2,E_4)$ will be Pareto optimal Nash equilibrium of the game if following condition holds true
\begin{eqnarray}
0.4655w_{P} \leq w_{I} \leq 4w_{1}+2w_{2}+w_{3}
\end{eqnarray} 
We now proceed to analyze the PP game for different choices of weights which may lead the game to different Nash equilibriums, which may or may not be Pareto optimal. We have specifically found Nash Equilibrium for following two cases:

\noindent \textbf{Case 1:} In order to study eavesdropping excluding DoS attacks, we consider $w_e=w_b=0$, which leads us to the set of Nash Equilibriums in the PP game, shown by Table VI \\
\textbf{Case 2:} For an Eve equipped with unlimited resources, we consider $w_1=w_2=w_3=0$, which leads us to the set of Nash Equilibriums in the PP game, shown by Table VII

\begin{table}[h!]
  \centering
  \caption{Conditions for $(A_{i}, E_{j})$ to be a Nash equilibrium for eavesdropping excluding DoS attacks}
  \label{tab:table6}
  \renewcommand{\arraystretch}{2}
  \begin{tabular}{|c|c|}   \hline
   Nash Equilibrium & Conditions \\  [1.0 ex]   \hline
    $(A_1,E_1)$ & \pbox{20cm} {$w_{d} \geq 1.2987w_{f}+25.974w_{1}+10.3896w_{2}+5.1948w_{3}$} \\ [1.0 ex] \hline
    $(A_1,E_4)$ & \pbox{20cm} {$w_{d} \leq 1.2987w_{f}+25.974w_{1}+10.3896w_{2}+5.1948w_{3}$} \\ [1.0 ex] \hline
    $(A_2,E_3)$ & \pbox{20cm} {$w_{d} \geq 4w_{1}+2w_{2}+w_{3}$} \\ [1.0 ex] \hline
    $(A_2,E_4)$ & \pbox{20cm} {$w_{d} \leq 4w_{1}+2w_{2}+w_{3}$} \\ [1.0 ex] \hline 
  \end{tabular}
\end{table} 
\begin{table}[h!]
  \centering
  \caption{Conditions for $(A_{i}, E_{j})$ to be a Nash equilibrium for an eavesdropper with unlimited power}
  \label{tab:table7}
  \renewcommand{\arraystretch}{2}
  \begin{tabular}{|c|c|}   \hline
   Nash Equilibrium & Conditions \\  [1.0 ex]   \hline
    $(A_1,E_1)$ & \pbox{20cm} {$w_{d} \geq 15.375w_{e}$, and \nonumber \\ $0.385w_{d}+0.689w_{e} \geq 0.5w_{f}$} \\ [2.0 ex] \hline
    $(A_1,E_2)$ & \pbox{20cm} {$w_{d} \leq 15.375w_{e}$, \nonumber \\ $0.377w_{d}+0.812w_{e} \geq 0.5w_{f}$, and \nonumber \\ $w_{b} \geq 0.601w_{a}$} \\ [2.0 ex] \hline
    $(A_1,E_3)$ & \pbox{20cm} {$0.385w_{d}+0.689w_{e} \leq 0.5w_{f}$, and \nonumber \\ $0.377w_{d}+0.812w_{e} \leq 0.5w_{f}$} \\ [2.0 ex] \hline
    $(A_1,E_4)$ & \pbox{20cm} {$0.385w_{d}+0.689w_{e} \leq 0.5w_{f}$, and \nonumber \\ $0.377w_{d}+0.812w_{e} \leq 0.5w_{f}$} \\ [2.0 ex] \hline
  \end{tabular}
\end{table} 

From VII, it is clear that for an eavesdropper with unlimited resources the condition for $(A_1,E_3)$ or $(A_1,E_4)$ to be Nash equilibrium is same which is justified as there are no costs involved for the resources to be used in eavesdropping. Therefore, the attack $(A_1,E_4)$ becomes irrelevant if Eve has unlimited power in terms of resources to be used. \par
Furthermore, an iterated version of the PPP can also be studied, so that knowledge of the previous moves of the opponents helps the players in deciding their next strategy. In the iterated version of the protocol, Alice will always prefer performing $A_1$, irrespective of what strategy Eve adopts in the previous step; and Eve may come to know that Alice always adopts $A_1$ and hence takes her move accordingly. Therefore, the Nash equilibrium of the game for an iterated protocol may only correspond to $A_1$ strategy of Alice. Apart from Eve slowly knowing the tendency of Alice adopting $A_1$, all other operations of Alice and Eve and the payoffs for the respective strategies remains same. 

We have seen the PPP game where payoffs of Alice and Eve are given by Eq. 2 and 3 respectively. We can also study the PPP game by modifying the payoffs to include Quantum Bit Error Rate (QBER) in Alice's and Eve's payoff. By doing so, we can include improvements introduced for making the protocol more secure \cite{Wojcik}.
 
\section{Comparison of PP protocol with LM05 protocol}
Standard QKD protocols such as BB84 protocol do not allow the receiver the decode the information in a deterministic way. This problem, however, can be rectified using a two-way QKD protocol such as PP protocol or LM05 protocol \cite{CaiLi2,LM05}. The LM05 protocol is based on nonorthogonal states instead of entangled resources as in PP protocol. In general, two-way QKD protocols have been proved better and secure against general eavesdropping attacks \cite{rev3a}. In this section, similar to the study of PP game, we design a generic two-way QKD game to analyze and compare PP and LMO5 protocols for some zero-loss eavesdropping attacks \cite{rev3b}. The payoffs of Alice and Eve in the general two-way QKD game can be described as 
\begin{eqnarray}
P_{A}=w_{g}I(A:B)-w_{h}[I(A:E)+I(B:E)]+w_{i}\left[ \dfrac{p_{d}+QBER}{2} \right] -w_jn
\end{eqnarray}
\begin{eqnarray}
P_{E} &=& w_{k}[I(A:E)+I(B:E)]-w_{l}I(A:B) + w_{m}\left[ 1-\dfrac{p_{d}+QBER}{2} \right]
\end{eqnarray}
where QBER is Quantum Bit Error Rate calculated by comprising some encoded bits (in message mode) shared between Alice and Bob; $n$ is the number of entangled states used; $w_g, w_h, w_i, w_j, w_k, w_l, w_m$ are positive real numbers and considered as weights attached to each quantity in the payoff. For various strategies of Eve, we study Intercept and Resend (IR) \cite{rev3b} attack, Double CNOT (DCNOT) \cite{rev3b} attack (which is also similar to Pavicic's attack \cite{Pavicic, Zawadzki}), and Wojcik's attack \cite{Wojcik}. In our present analysis, we consider PPP with encoding scheme $A_1$ as described above, and LM05 \cite{LM05} protocol. The payoffs of Alice and Eve for PPP and LM05 during various eavesdropping attacks are summarized in Table VIII and Table IX.

\vspace*{-2mm}
\begin{table}[h!]
  \centering
  \caption{Payoffs of Alice in the two-way QKD game}
  \label{tab:table8}
  \renewcommand{\arraystretch}{2}
  \begin{tabular}{|c|c|c|c|}   \hline
    \backslashbox{Alice}{Eve} & $IR$ & $DCNOT$ & $Wojcik'sAttack$ $(E_1)$ \\  [1.0 ex]   \hline
    $PPP$ & \pbox{20cm} {$0.1887w_{g}-1.1887w_{h}+0.125w_{i}-w_{j}$} & \pbox{20cm} {$w_{g}-w_{j}$} & \pbox{20cm} {$0.311w_{g}-0.385w_{h}+0.375w_{i}-w_{j}$} \\ [2.0 ex]     \hline
    $LM05$ & \pbox{20cm} {$0.1887w_{g}-1.1887w_{h}+0.125w_{i}$} & \pbox{20cm} {$w_{g}-2w_{h}+0.125w_{i}$} & \pbox{20cm} {$0.5488w_{g}-1.096w_{h}+0.375w_{i}$} \\ [2.0 ex]     \hline
  \end{tabular}
\end{table}
\vspace*{2mm}
\vspace*{-2mm}
\begin{table}[h!]
  \centering
  \caption{Payoffs of Eve in the two-way QKD game}
  \label{tab:table9}
  \renewcommand{\arraystretch}{2}
  \begin{tabular}{|c|c|c|c|}   \hline
    \backslashbox{Alice}{Eve} & $IR$ & $DCNOT$ & $Wojcik'sAttack$ $(E_1)$ \\  [1.0 ex]   \hline
    $PPP$ & \pbox{20cm} {$1.1887w_{k}-0.1887w_{l}+0.875w_{m}$} & \pbox{20cm} {$-w_{l}+w_{m}$} & \pbox{20cm} {$0.385w_{k}-0.311w_{l}+0.625w_{m}$} \\ [2.0 ex]     \hline
    $LM05$ & \pbox{20cm} {$1.1887w_{k}-0.1887w_{l}+0.875w_{m}$} & \pbox{20cm} {$2w_{k}-w_{l}+0.875w_{m}$} & \pbox{20cm} {$1.096w_{k}-0.5488w_{l}+0.625w_{m}$} \\ [2.0 ex]     \hline
  \end{tabular}
\end{table}
\vspace*{2mm}
\par Table VIII clearly shows that LM05 game results in better payoff for Alice in comparison to the PP game, for the IR attack. However, if Eve performs DCNOT or Wojcik's attack, Alice may get better payoff by playing either PP game or LM05 game depending on the values of the weights $w_g$, $w_h$, $w_i$, and $w_j$. Similarly, Table IX suggests that Eve will prefer IR attack over Wojcik's attack for both the games. Moreover, Eve may prefer either IR or DCNOT attack depending on the values of weights $w_k$, $w_l$, and $w_m$. Hence, Nash equilibrium of the two-way QKD game varies for different conditions of weights as shown in Table X. 
\begin{table}[h!]
  \centering
  \caption{Conditions for $(A_{i}, E_{j})$ to be a Nash equilibrium in the two-way QKD game}
  \label{tab:table10}
  \renewcommand{\arraystretch}{2}
  \begin{tabular}{|c|c|}   \hline
   Nash Equilibrium & Conditions \\  [1.0 ex]   \hline 
    $(LM05,IR)$ & \pbox{20cm} {$w_{k} \leq w_{l}$} \\ [2.0 ex] \hline
    $(PPP,DCNOT)$ & \pbox{20cm} {$2w_{h}-w_{j} \geq 0.125w_{i}$ \\ $w_{m} \geq 9.5096w_{k}+6.4904w_{l}$}	\\ [2.0 ex] \hline
    $(LM05,DCNOT)$ & \pbox{20cm} {$2w_{h}-w_{j} \leq 0.125w_{i}$ \\ $w_{k} \geq w_{l}$ \\ $w_{m} \geq 1.8048w_{l}-3.616w_{k}$} \\ [2.0 ex] \hline
  \end{tabular}
\end{table} 

\section{Conclusion}
We have analysed PP protocol from the point of view of a game with quantum strategies. Our results established a relation between pure strategy Nash equilibrium, and payoffs of the sender and the eavesdropper depending on the value of weights assigned to mutual information between different players, probability of detection of the eavesdropper and number of gates applied by an eavesdropper to gain information. We have shown the strategy that a sender must opt to minimize the payoff of an eavesdropper within the conditions of the game described in this article. Our analysis further demonstrated the condition for a Pareto optimal Nash equilibrium. In order to study a general two-way QKD protocol with and without entanglement, we compared the PP protocol with LM05 protocol. We found that the payoffs of the sender and eavesdropper depend on the eavesdropping attacks and the weights of different terms playing crucial role in the designed payoff. We believe that the study presented here will provide a deeper understanding of PP protocol from the perspective of strategies employed by a sender or an eavesdropper to achieve a better payoff.

\end{document}